\documentclass[
11pt,
reprint,
onecolumn,
tightenlines,
superscriptaddress,
preprintnumbers,
nofootinbib,
amsmath,amssymb,amsthm,
physrev,
eqsecnum,tikz,
]{revtex4-2}

\usepackage{isomath}
\usepackage{amsmath,amsthm}
\usepackage{amsbsy}
\usepackage{amssymb}
\usepackage{amscd}
\usepackage{amsfonts}
\usepackage{stmaryrd}
\usepackage{siunitx}
\usepackage{euscript}
\usepackage[utf8]{inputenc}
\usepackage[T1]{fontenc}
\usepackage{newtxtext} 
\everymath{\displaystyle}
\usepackage{exscale}
\usepackage{microtype}
\usepackage{hyperref}
\usepackage{booktabs}
\usepackage{algorithm}
\usepackage{algpseudocode}

\usepackage{graphicx}
\usepackage{boxedminipage}
\usepackage{calc}
\usepackage[usenames,dvipsnames]{xcolor}
\graphicspath{ {media/} }
\usepackage[caption=false,justification=centerlast]{subfig}

\usepackage{setspace}
\usepackage{enumitem}
\setitemize{noitemsep,topsep=0pt,parsep=0pt,partopsep=0pt}
\setenumerate{noitemsep,topsep=0pt,parsep=0pt,partopsep=0pt}
\setdescription{noitemsep,topsep=0pt,parsep=0pt,partopsep=0pt}

\usepackage[colorinlistoftodos, color=green!40,prependcaption]{todonotes}
\setuptodonotes{inline}

\usepackage{soul} % highlighting
\usepackage[normalem]{ulem}

\usepackage{orcidlink}
\usepackage{siunitx}

\usepackage[small]{titlesec}

\titlespacing*{\section}{0pt}{12pt plus 4pt minus 2pt}{2pt plus 2pt minus 2pt}
\titlespacing*{\subsection}{0pt}{12pt plus 4pt minus 2pt}{2pt plus 2pt minus 2pt}
\titlespacing*{\subsubsection}{0pt}{12pt plus 4pt minus 2pt}{2pt plus 2pt minus 2pt}
\titlespacing*{\paragraph}{0pt}{12pt plus 4pt minus 2pt}{2pt plus 2pt minus 2pt}

\makeatletter

    \renewcommand*{\p@subsection}{}
    
    \renewcommand*{\p@subsubsection}{}
\makeatother

\usepackage{isomath}
\usepackage{amsmath}
\usepackage{amssymb}
\usepackage{amscd}
\usepackage{amsfonts}
\usepackage{upgreek}

\newcommand{\half}{\tfrac{1}{2}}

\theoremstyle{definition}

\AtEndEnvironment{definition}{\null\hfill\qedsymbol}
\newtheorem{remark}{Remark}[section]
\AtEndEnvironment{remark}{\null\hfill\qedsymbol}

\AtEndEnvironment{example}{\null\hfill\qedsymbol}

\AtEndEnvironment{assumption}{\null\hfill\qedsymbol}

\newcommand{\bfsigma}{\mathbold {\sigma}}

\DeclareMathOperator{\divergence}{div}

\newcommand{\parderiv}[2]{\frac{\partial #1}{\partial #2}}
\newcommand{\dm}{\ \mathrm{d}}

\newcommand{\bfa}{{\mathbold a}}

\newcommand{\bfe}{{\mathbold e}}

\newcommand{\bfn}{{\mathbold n}}

\newcommand{\bfu}{{\mathbold u}}

\newcommand{\bfx}{{\mathbold x}}

\newcommand{\bfC}{{\mathbold C}}
\newcommand{\bfD}{{\mathbold D}}
\newcommand{\bfE}{{\mathbold E}}
\newcommand{\bfF}{{\mathbold F}}

\newcommand{\bfI}{{\mathbold I}}

\newcommand{\bfL}{{\mathbold L}}

\newcommand{\bfP}{{\mathbold P}}

\newcommand{\bfU}{{\mathbold U}}

\newcommand{\bfeps}{\mathbold{\varepsilon}}

\begin{document}

\preprint{To appear in Journal of Applied Mechanics (DOI: \href{ https://doi.org/10.1115/1.4066285}{10.1115/1.4066285})}

\title{Interplay between Nucleation and Kinetics in Dynamic Twinning}

\author{Janel Chua}
    \email{janelchua@lanl.gov}
    \affiliation{Los Alamos National Laboratory}

\author{Vaibhav Agrawal}
    \affiliation{Apple Inc.}
    
\author{Noel Walkington}
    \affiliation{Center for Nonlinear Analysis, Department of Mathematical Sciences, Carnegie Mellon University}

\author{George Gazonas}
    \affiliation{DEVCOM Army Research Laboratory, Attn: FCDD-RLW-MB, Aberdeen Proving Ground, MD 21005, USA}

\author{Kaushik Dayal}
    \affiliation{Department of Civil and Environmental Engineering, Carnegie Mellon University}
    \affiliation{Center for Nonlinear Analysis, Department of Mathematical Sciences, Carnegie Mellon University}
    \affiliation{Department of Mechanical Engineering, Carnegie Mellon University}
    
\date{\today}

%%%%%%%%%%%%%%%%%%%%%
%%%%%%%%%%%%%%%%%%%%%
%%%%%%%%%%%%%%%%%%%%%
%%%%%%%%%%%%%%%%%%%%%

\begin{abstract}

    In this work, we apply a phase-field modeling framework to elucidate the interplay between nucleation and kinetics in the dynamic evolution of twinning interfaces.
    The key feature of this phase-field approach is the ability to transparently and explicitly specify nucleation and kinetic behavior in the model, in contrast to other regularized interface models.
    We use this to study 2 distinct problems where it is essential to explicitly specify the kinetic and nucleation behavior governing twin evolution.

    First, we study twinning interfaces in 2-d.
    When these interfaces are driven to move, we find that significant levels of twin nucleation occur ahead of the moving interface.
    Essentially, the finite interface velocity and the relaxation time of the stresses ahead of the interface allows for nucleation to occur before the interface is able to propagate to that point.
    Second, we study the growth of needle twins in antiplane elasticity.
    We show that both nucleation and anisotropic kinetics are essential to obtain predictions of needle twins.
    While standard regularized interface approaches do not permit the transparent specification of anisotropic kinetics, this is readily possible with the phase-field approach that we have used here.

\end{abstract}

\maketitle

%%%%%%%%%%%%%%%%%%%%%
%%%%%%%%%%%%%%%%%%%%%
%%%%%%%%%%%%%%%%%%%%%
%%%%%%%%%%%%%%%%%%%%%
\section{Introduction}

Twinning and structural phase transformations are important phenomena in materials science and solid-state physics due to the significant role they play in influencing the properties and behavior of materials \cite{porter2009phase}.
Some specific areas of applications are in shape memory alloys (SMAs) where the reversible twinning and detwinning of SMAs are responsible for their unique shape-changing behavior \cite{abeyaratne2006evolution}.
In other instances, structural phase transformations allow for tailoring material properties to specific applications, for instance in nanotwinned metals \cite{jiao2015molecular,jiao2018radiation,sinha2014alternating,kulkarni2009some}.

The microstructure of the materials in which twinning or phase transformations occur typically consists of homogeneously deformed regions separated by interfaces across which the deformation varies extremely rapidly.
Many interesting material phenomena under extreme conditions are governed by the nucleation, motion and response of the interfaces \cite{faran2011kinetic,abeyaratne2006evolution}.
It is therefore important to be able to accurately model propagating interfaces in a dynamic setting, particularly when inertia plays a significant role since these interfaces move at velocities comparable to the sonic speeds.

\paragraph*{Related prior work.}

Continuum models for dynamic twin interface nucleation and propagation fall into two broad classes: sharp interface models \cite{abeyaratne1990driving,abeyaratne1991kinetic,truskinovskii1982equilibrium,truskinovsky1993kinks,rosakis-knowles-non-monotone,Rosakis-levelset} and regularized interface models \cite{abeyaratne1991implications, rosakis1995equal,turteltaub1997viscosity,fried1994dynamic,yang2010formulation,knap-clayton}. 
In the case of the former, numerical computations are extremely challenging as sharp interfaces require expensive tracking algorithm in a numerical discretization. 
This is essentially infeasible when one expects numerous evolving and interacting interfaces, and thus the sharp interface approaches are not widely applied. 

Regularized interfaces on the other hand do not require explicit tracking of the interfaces, which makes the numerical methods much simpler and enables application to complex settings.
Phase-field models, wherein an additional regularized phase field variable is introduced to account for material transformations, have been widely applied to problems such as structural phase transformations, fracture, poromechanics, and dislocation dynamics, e.g. \cite{karma-crack-2001,clayton2014geometrically,penrose1990thermodynamically,hakimzadeh2022phase,agrawal2017dependence,dayal2007real,chua2024deformation,karimi2022energetic,karimi2023high,Clayton2023,Clayton2024}.
However, these methods also have significant shortcomings.
Particularly, the dependence of the kinetics and nucleation of interfaces on the structure of the material model is completely opaque.
Further, the range of material responses that can be obtained is highly constrained. 
Furthermore, the nucleation and the kinetics of interfaces are physically distinct processes from the atomic perspective, but standard phase-field models are unable to transparently distinguish these processes.
One consequence of these shortcomings is that even in the simplest 1-d case, the precise critical condition at which nucleation occurs is completely opaque. 
Consequently, it is essentially impossible in practice to formulate a model to obtain a desired nucleation response.
All of this is in sharp contrast to sharp interface models wherein the nucleation and kinetics of interfaces can be transparently specified in the model.

To retain the computational ease of phase-field models and at the same time obtaining the transparent physical structure of sharp-interface models, we recently proposed a phase-field modeling approach that enables the transparent specification of interface nucleation and kinetics as input to the model \cite{agrawal2015dynamic,agrawal2015dynamic-2,chua2022phase}.
The key elements of this approach are, first, formulating an energy density that separates nucleation from kinetics; and, second, formulating a geometrically-motivated interface conservation principle to govern the nucleation and kinetics of interface. 
In this conservation principle, the kinetics of interfaces is governed by a transport term and the nucleation of interfaces is governed by a source term, and each of these contribution can be transparently and independently specified.

\paragraph*{Contributions of this paper.}

In this paper, we use the phase-field approach described above to study the interplay between kinetics and nucleation in dynamic twinning.
Specifically, we study 2 distinct problems where it is essential to be able to explicitly specify the kinetic and nucleation behavior governing twin evolution.

First, we study twinning interfaces in 2-d.
When these interfaces are driven to move, we find that significant levels of twin nucleation occur ahead of the moving interface.
Essentially, the finite interface velocity and the relaxation time of the stresses ahead of the interface allows for nucleation to occur before the interface is able to propagate to that point.
Second, we study the growth of needle twins in antiplane elasticity.
We show that both nucleation and anisotropic kinetics are essential to obtain predictions of needle twins.
While standard regularized interface approaches do not permit the transparent specification of anisotropic kinetics, this is readily possible with the phase-field approach that we have used here.

%%%%%%%%%%%%%%%%%%%%%
%%%%%%%%%%%%%%%%%%%%%
%%%%%%%%%%%%%%%%%%%%%
%%%%%%%%%%%%%%%%%%%%%
\section{Model Formulation}
\label{sec:formulation}

%%%%%%%%%%%%%%%%%%%%%
%%%%%%%%%%%%%%%%%%%%%
%%%%%%%%%%%%%%%%%%%%%
%%%%%%%%%%%%%%%%%%%%%
\subsection{Notation}

We use $\bfx_0$ and $\bfx(\bfx_0)$ to denote the reference and deformed configurations, and the displacement is denoted $\bfu(\bfx_0) := \bfx(\bfx_0) - \bfx$.
We work completely in the Lagrangian setting for simplicity, while noting that there can be several advantages to an Eulerian approach in the large deformation setting \cite{naghibzadeh2021surface,naghibzadeh2022accretion,Freed2020,Clayton2013}.
We use $\nabla$ and $\divergence$ to denote the gradient and divergence with respect to $\bfx_0$, and superposed dots to denote time derivatives.
The deformation gradient is denoted by $\bfF := \nabla \bfx$, the strain by  $\bfE := \half\left(\bfF^T \bfF - \bfI\right)$, the spatial velocity gradient by $\bfL = \dot{\bfF}\bfF^{-1}$, and the first Piola-Kirchoff stress by $\bfP$.
We define $\bfD := \half\left(\bfL + \bfL^T\right)$, $J := \det \bfF$, and use $\phi$ to denote the phase field.

We define a regularized indicator function $H_l(\cdot)$ that smoothly transitions --- over an interval of order $l$ --- from $0$ to $1$ when the argument changes from negative to positive.
For specificity, we use the form $H_l\left(x\right) = \half\left(1+\tanh\left(\frac{x}{l}\right)\right)$ and $l=0.1$.

The domain and its boundary are denoted by $\Omega$ and $\partial\Omega$ respectively.

%%%%%%%%%%%%%%%%%%%%%
%%%%%%%%%%%%%%%%%%%%%
%%%%%%%%%%%%%%%%%%%%%
%%%%%%%%%%%%%%%%%%%%%
\subsection{Governing Equations}

We use the standard free energy of the form:
\begin{equation}
    E[\bfu,\phi] = \int_\Omega \left(W(\bfF,\phi) + \frac{\alpha}{2} |\nabla\phi|^2\right) \dm\Omega
    \label{consolidation problem energy functional}
\end{equation}
where $W$ is the elastic energy density and $\alpha$ is the phase-field regularization parameter.
We use $\alpha=\num{1.2e-4}\unit{\meter}$.

We follow \cite{agrawal2015dynamic,agrawal2015dynamic-2} in setting $W$ to have the general form:
\begin{equation}
    W(\bfF,\phi) 
    = 
    H_l\left( 0.5 - \phi \right) W_1(\bfF)
    + 
    H_l\left( \phi-0.5 \right) W_2(\bfF)
    \label{consolidation problem energy functional_strain energy density}
\end{equation}
where $W_1$ and $W_2$ correspond to the elastic energy density in phase 1 ($\phi < 0.5$) and phase 2 ($\phi > 0.5$) respectively.
As described in \cite{agrawal2015dynamic,agrawal2015dynamic-2}, this form contributes to a transparent separation between kinetics and nucleation.    
For specificity, we choose $W_i(\bfF) = \half \left(\bfE-\bfE_i\right) : \bfC_i : \left(\bfE-\bfE_i\right), i \in \{1,2\}$, where $\bfC_i$ is the 4-th order elastic modulus tensor.

The evolution equations associated with this energy are \cite{agrawal2015dynamic,agrawal2015dynamic-2}:
\begin{subequations}
\begin{align}
    &\text{Momentum balance: } \rho \ddot{\bfu} = \divergence\bfP
    \label{eqn:consolidation disp-form}
    \\
    &\text{Phase evolution: } \dot{\phi} = |\nabla\phi|\hat{v}\left(f\right) + G(\bfE, \phi)
    \label{eqn:consolidation phi-form}
\end{align}
\end{subequations}

The evolution equation for $\phi$ was proposed in \cite{alber2005solutions}, and was shown to correspond to a balance law for interfaces \cite{agrawal2015dynamic-2,guin2023phase}.
$G$ originates as a source term in that balance law, and in this work is used to induce the nucleation of an appropriate phase when the system evolves to access nonphysical regions of the energy landscape \cite{chua2022phase}.
The specific forms of $G$ will be described in later sections.

The phase velocity $\hat{v}(f)$ is a function of the driving force $f$, which is the negative of the functional derivative of $E$ with respect to $\phi$:
\begin{equation}
    f := -\parderiv{W}{\phi} + \alpha\divergence\nabla\phi
    \label{eqn:consolidation f-form}
\end{equation}

\cite{agrawal2015dynamic,agrawal2015dynamic-2,chua2022phase} provide conditions that $G$ and $\hat{v}(f)$ must satisfy for consistency with the 2-nd law of thermodynamics \cite{karimi2022energetic,penrose1990thermodynamically}.

The Piola-Kirchoff stress is composed of an elastic part and a viscous dissipative part: $\bfP = \bfP_{elas} + \bfP_{vis}$.
The elastic part is given by $\bfP_{elas} = \left(\parderiv{W(\bfF,\phi)}{\bfF}\right)$.
The viscous stress is assumed to follow a frame-invariant Newtonian model with a linear relation between the Cauchy stress and $\bfD$ \cite{sengul2021nonlinear}:
\begin{equation}
    \bfsigma_{vis} 
    = 
    \eta\bfD 
    = 
    \frac{\eta}{2}\left(\bfL + \bfL^T\right) 
    \implies
    \bfP_{vis} 
    = 
    J\bfsigma_{vis}\bfF^{-T}
    =
    \frac{\eta}{2} J \left(\Dot{\bfF}\bfF^{-1} + \bfF^{-T}\Dot{\bfF}^{T}\right) \bfF^{-T}
\end{equation}
where $\eta$ is the viscosity.

%%%%%%%%%%%%%%%%%%%%%
%%%%%%%%%%%%%%%%%%%%%
%%%%%%%%%%%%%%%%%%%%%
%%%%%%%%%%%%%%%%%%%%%
\subsection{Finite Element Implementation}

Weuse the finite element method (FEM) to solve numerically using the using the open-source FEM package FEniCS \cite{Fenics,Fenics2}.
We use a mixed finite element formulation and discretize \eqref{eqn:consolidation disp-form}, \eqref{eqn:consolidation f-form}, and \eqref{eqn:consolidation phi-form}.
The mixed formulation is required because the higher derivatives in $f$ from \eqref{eqn:consolidation f-form} cannot be treated in the usual way using integration-by-parts, because $f$ appears inside the response function $\hat{v}(f)$.
The weak form is then:
\begin{subequations}
\begin{align}
    \int_\Omega \rho\ddot{\bfu}\cdot \tilde{\bfu} \dm\Omega
    & =
    -\int_\Omega \bfP:\nabla\tilde{\bfu} \dm\Omega + \int_{\partial\Omega} (\bfP \hat{\bfn})\cdot \tilde{\bfu} \dm S 
    \label{eqn:consolidation weak-form-disp}
    \\
    \int_\Omega f\tilde{f}\dm\Omega 
    & = 
    \int_\Omega \left(-\left(\parderiv{W}{\phi}\right)\tilde{f} - \alpha\nabla\phi\cdot\nabla\tilde{f} \right) \dm\Omega 
    + 
    \int_{\partial\Omega} \alpha\tilde{f}\nabla\phi\cdot\hat{\bfn} \dm S
    \label{eqn:consolidation weak-form-f}
    \\
    \int_\Omega \dot{\phi}\tilde{\phi}\dm\Omega
    & =  
    \int_\Omega \left( |\nabla \phi|\hat{v}(f) + G(\bfE, \phi)\right) \tilde{\phi}\dm\Omega 
    \label{eqn:consolidation weak-form-phi}
\end{align}
\label{consolidation weak-form}
\end{subequations}
where $\bfu$ and the corresponding test function $\tilde{\bfu}$ belong to the space of vector-valued functions that are piecewise linear; and $f$ and $\phi$, with the corresponding test functions $\tilde{f}$ and $\tilde{\phi}$, belong to the space of scalar-valued functions that are piecewise linear.

The boundary conditions for momentum balance \eqref{eqn:consolidation weak-form-disp} are the standard displacement or traction prescribed, and the boundary condition for phase evolution \eqref{eqn:consolidation weak-form-phi} is the standard variational condition $\nabla\phi\cdot\hat{\bfn} = 0$ \cite{yang2010formulation} on the entire boundary.

The momentum balance and the phase evolution are evolved in time using the implicit Newmark method and the implicit Euler method respectively.

%%%%%%%%%%%%%%%%%%%%%
%%%%%%%%%%%%%%%%%%%%%
%%%%%%%%%%%%%%%%%%%%%
%%%%%%%%%%%%%%%%%%%%%
\section{Nucleation Ahead of the Twin Interface at High Rates}

We apply the model described in Section \ref{sec:formulation} to predict the motion of a flat twin interface in a 2-d nonlinear elastic setting, and specifically study the role of phase nucleation when inertial effects are significant.

%%%%%%%%%%%%%%%%%%%%%
%%%%%%%%%%%%%%%%%%%%%
%%%%%%%%%%%%%%%%%%%%%
%%%%%%%%%%%%%%%%%%%%%
\subsection{Elasticity and Geometry of Twinning}

Following \cite{agrawal2015dynamic-2}, we consider twinning defined by the following transformation stretch tensors:
%\begin{subequations}
\begin{align}
    &\bfU_1 = 
    \begin{bmatrix}
        a & 0 \\
        0 & b 
   \end{bmatrix}, 
   \quad 
   \bfU_2 = 
   \begin{bmatrix}
        b & 0 \\
        0 & a 
   \end{bmatrix}, 
   \quad a = 0.8958, \quad b = 1.09659 \label{twin stretch tensors} 
\end{align}
The stress-free strains are defined by $\bfE_1 = \frac{1}{2}\left(\bfU_1^2 - \bfI\right)$ and $\bfE_2 = \frac{1}{2}\left(\bfU_2^2 - \bfI\right)$. 
This twinning system admits stress-free compatible twinning interfaces with (referential) normal given by $\frac{1}{\sqrt{2}} \left( \bfe_1 \pm \bfe_2\right)$ where $\bfe_1, \bfe_2$ are the unit vectors in the Cartesian directions \cite{Dayal2006,clayton-book,abeyaratne2006evolution}.

We assume that the elasticity tensor $\bfC$ is isotropic and identical in both phases, and set the extensional modulus to $\num{206}\unit{\giga\pascal}$ and the Poisson ratio to $0.3$.

\begin{figure*}[htb!]
    \centering
    \includegraphics[scale=0.6]{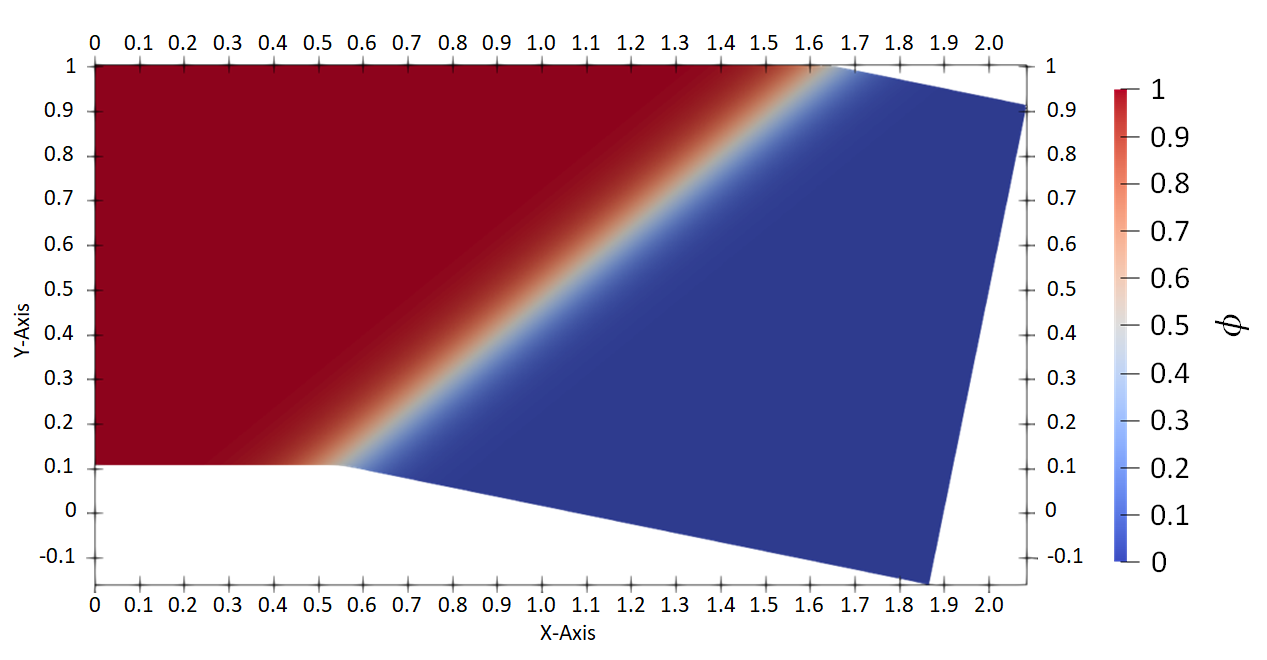}
    \caption{The deformed configuration of a rectangular block of size $2\times 1$ with $\phi$ overlaid, after energy minimization with traction-free boundary conditions all around except for the horizontal displacement set to $0$ on the left edge.}
    \label{U_Phi minimized}
\end{figure*}

To set up the initial condition for dynamic calculations described later on, we obtain a stress-free compatible interfaces by setting $\phi=1$ on the right of the domain and $\phi=0$ on the left side, and then minimize the energy in \eqref{consolidation problem energy functional}. 
The resulting static twin interface is shown in Figure \ref{U_Phi minimized}.

%%%%%%%%%%%%%%%%%%%%%
%%%%%%%%%%%%%%%%%%%%%
%%%%%%%%%%%%%%%%%%%%%
%%%%%%%%%%%%%%%%%%%%%
\subsection{Nucleation Model}
\label{physically plausible G term and vis stress}

\begin{figure*}[htb!]%
    \centering{\includegraphics[scale=0.67]{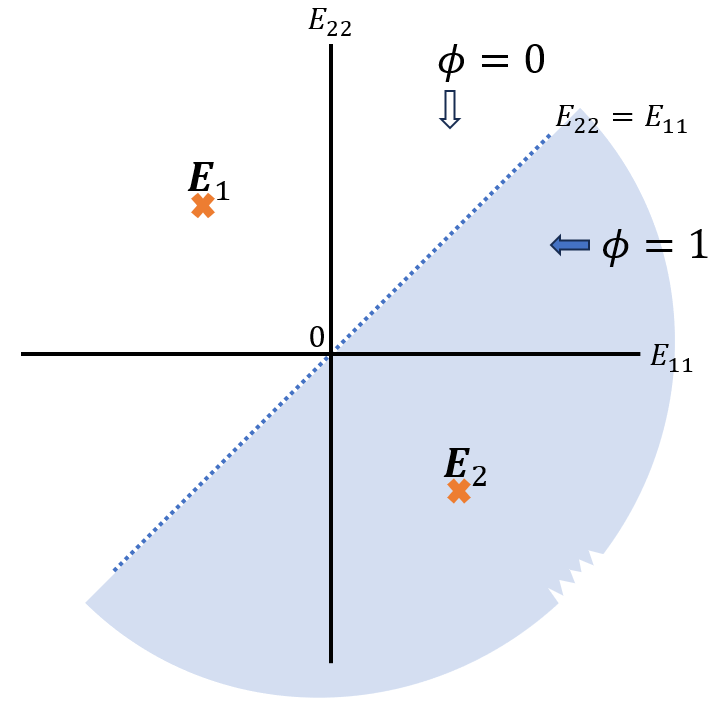}}
    \caption{Schematic depiction of the correspondence between the strain $\bfE$ and the phase $\phi$.
        We need $\phi < 0.5$ when $E_{22} > E_{11}$ and $\phi>0.5$ when $E_{22} < E_{11}$.}
    \label{consolidation G_nucleation illustration}
\end{figure*}

We first find the hyperplane in strain space that separates the twinning phases.
Considering strains corresponding to the stretch tensors given in \eqref{twin stretch tensors}, the hyperplane that is the set of points that are of equal distance from $\bfE_1$ and $\bfE_2$ has the equation $|\bfE^* - \bfE_2| = |\bfE^* - \bfE_1|$.
Solving for $\bfE^*$ gives $E^*_{11} = E^*_{22}$ as the equation of the hyperplane.
On one side of the hyperplane, we are closer to $\bfE_1$ and $\phi$ should be less than $0.5$; on the other side, we are closer to $\bfE_2$ and require that $\phi$ is greater than $0.5$ (Fig. \ref{consolidation G_nucleation illustration}).

We set up the nucleation criterion to enforce this correspondence, i.e., if we cross the hyperplane in strain space, we ensure that $\phi$ evolves appropriately.
Specifically, the nucleation process must drive $\phi$ to below $0.5$ when $E_{22}-E_{11}>0$, and drive $\phi$ to above $0.5$ when $E_{22}-E_{11}<0$, which results in constructing a nucleation term of the form:
\begin{align}
    &G(\bfE, \phi) = G_0\big( H_l\left(0.5-\phi\right) H_l\left(E_{11} - E_{22}\right) - H_l\left(\phi-0.5\right)H_l\left(E_{22} - E_{11}\right) \big)
\label{consolidation proper G term}
\end{align}

Since the role of kinetics is not the central focus here, we use the simple linear form $\hat{v}\left(f\right) = \kappa f$.
We use $\kappa=1$ for the numerical calculations.

%%%%%%%%%%%%%%%%%%%%%
%%%%%%%%%%%%%%%%%%%%%
%%%%%%%%%%%%%%%%%%%%%
%%%%%%%%%%%%%%%%%%%%%
\subsection{Results}
\label{high velocity results consolidation problem}

\begin{figure*}[htb!]%
    \centering{\includegraphics[scale=0.5]{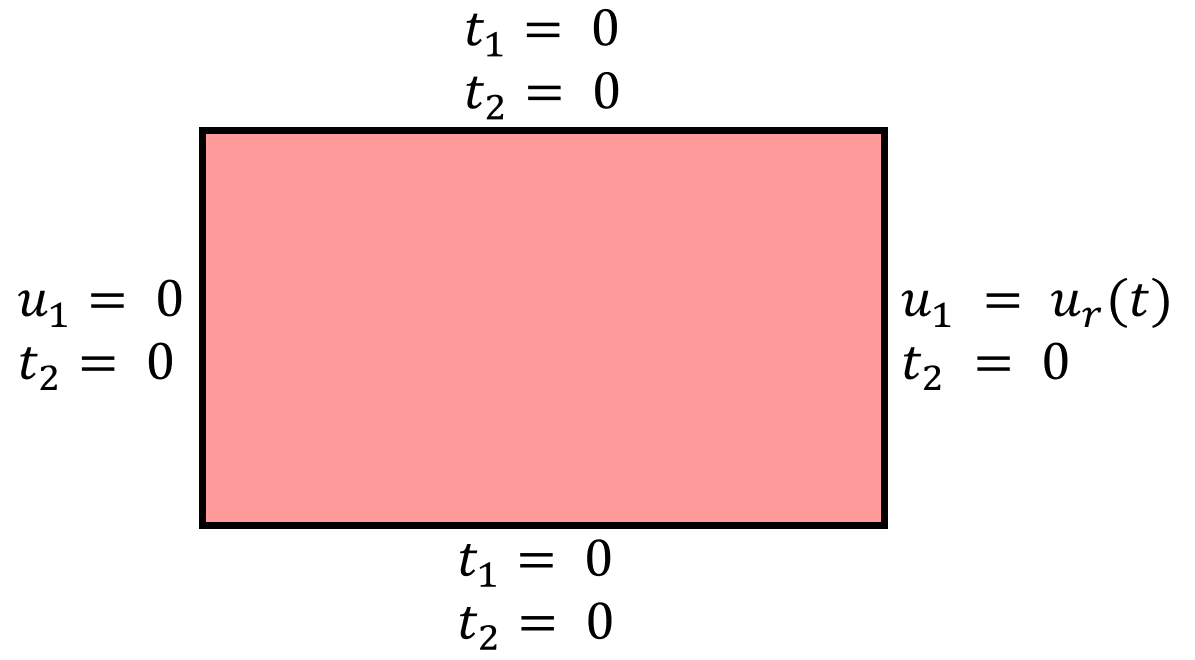}}
    \caption{Boundary conditions, where $t_1, t_2$ refer to Cartesian components of traction, and $u_1, u_2$ refer to Cartesian components of the displacement. The prescribed time-dependent loading is $u_r(t)$.}\label{consolidation boundary condition}
\end{figure*}

We perform dynamic calculations using the boundary conditions shown in Figure \ref{consolidation boundary condition}.
The boundary conditions are set up to minimally constrain the deformation of the body, and this allows a rigid vertical translation mode, but this mode does not interfere with our observation of the twinning deformation.
We use the static equilibrium interface as the initial conditions (Fig. \ref{U_Phi minimized}).

\begin{figure*}[htb!]
    \centering
    \subfloat[$u_r=0\unit{mm}$, $G_0=0\unit{s^{-1}}$]{\includegraphics[scale=0.39]{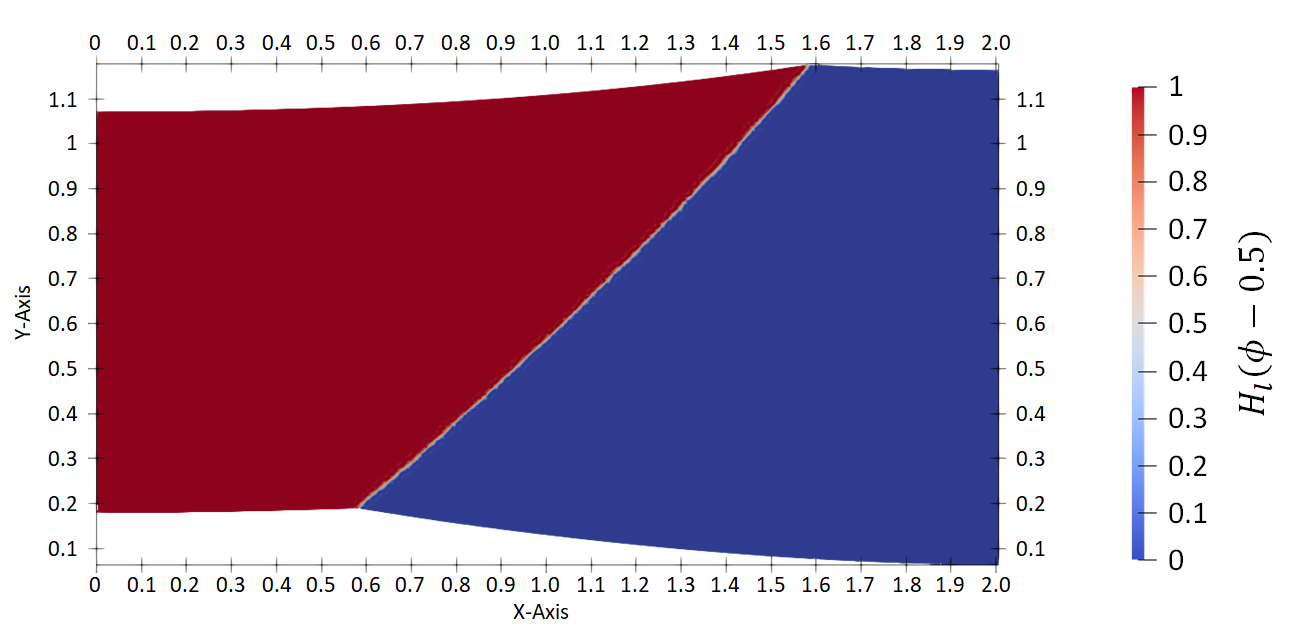}}
    \hfill
    \subfloat[$u_r=0\unit{mm}$, $G_0=\num{1e5}\unit{s^{-1}}$]{\includegraphics[scale=0.39]{media/Figure4.png}}
    \\
    \subfloat[$u_r=0.08\unit{mm}$, $G_0=0\unit{s^{-1}}$]{\includegraphics[scale=0.39]{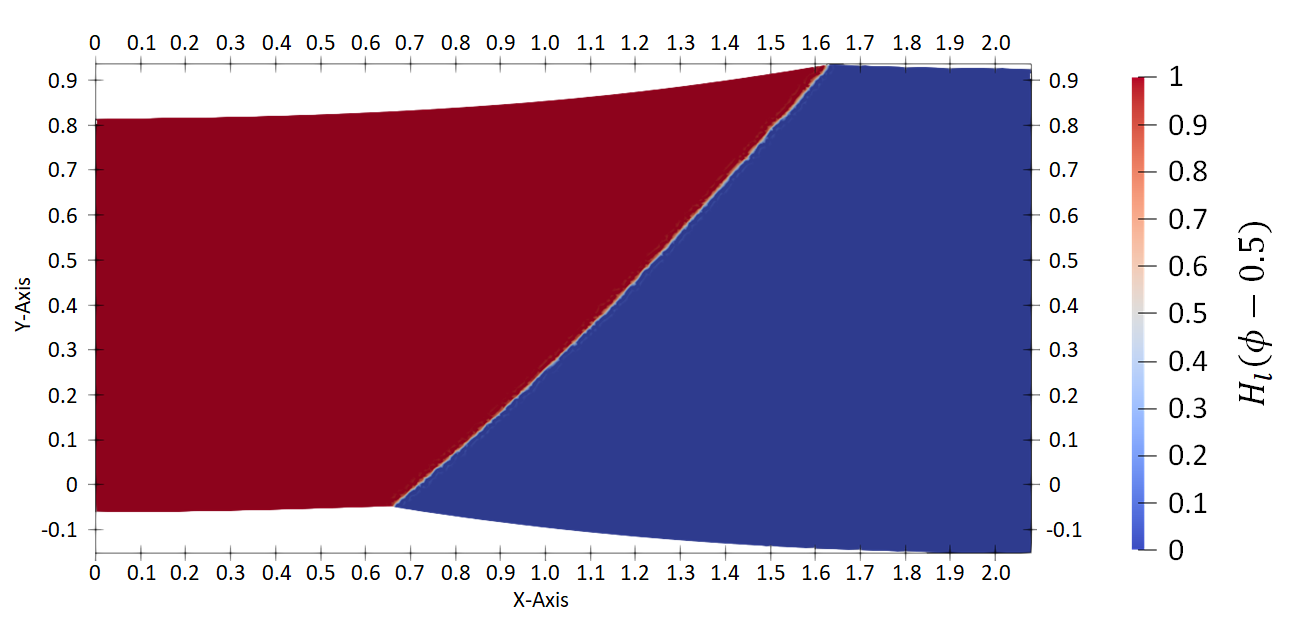}}
    \hfill
    \subfloat[$u_r=0.08\unit{mm}$, $G_0=\num{1e5}\unit{s^{-1}}$]{\includegraphics[scale=0.39]{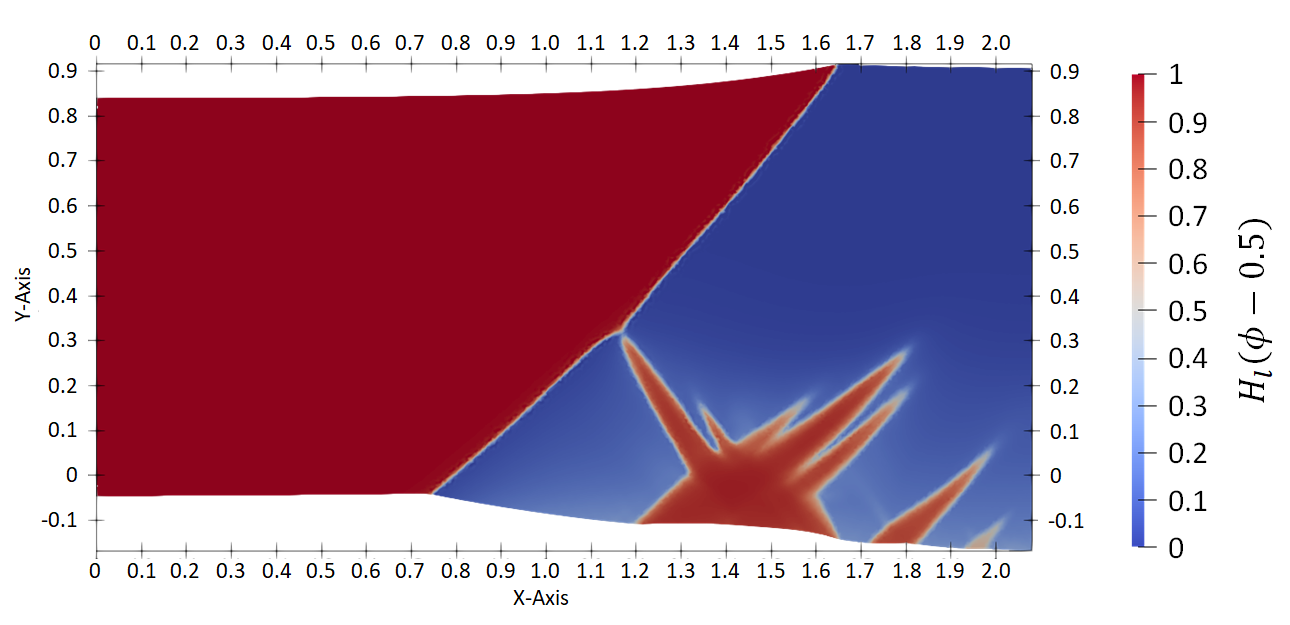}}
    \\
    \subfloat[$u_r=0.12\unit{mm}$, $G_0=0\unit{s^{-1}}$]{\includegraphics[scale=0.39]{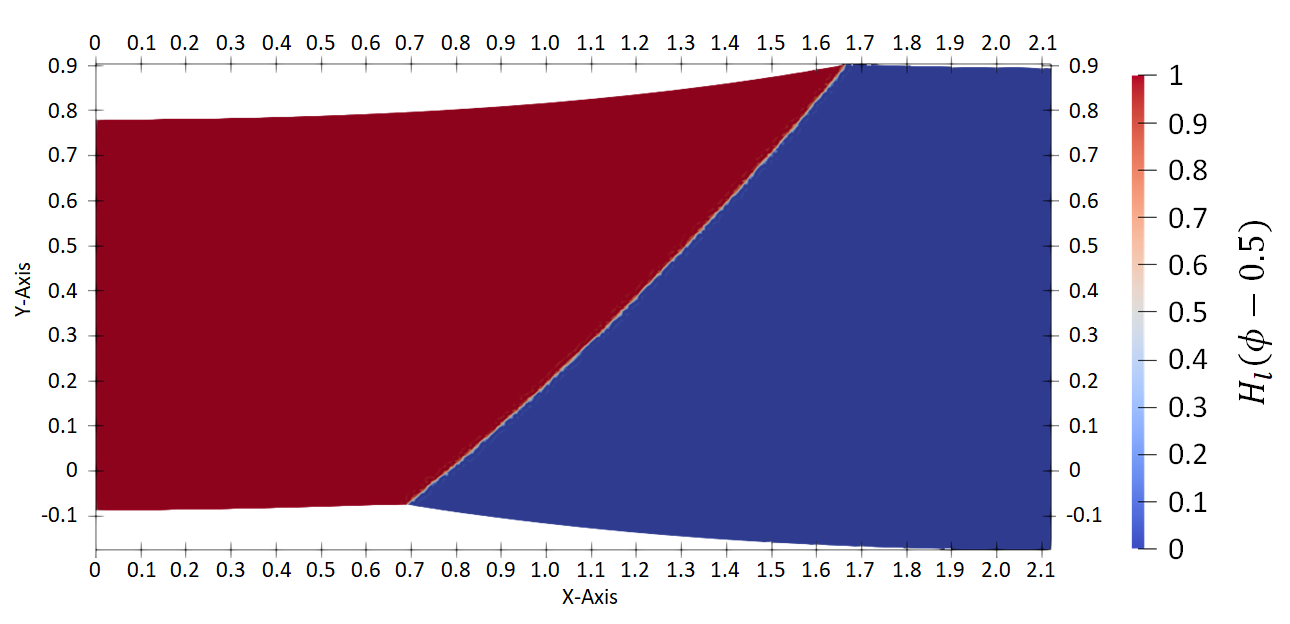}}
    \hfill
    \subfloat[$u_r=0.12\unit{mm}$, $G_0=\num{1e5}\unit{s^{-1}}$]{\includegraphics[scale=0.39]{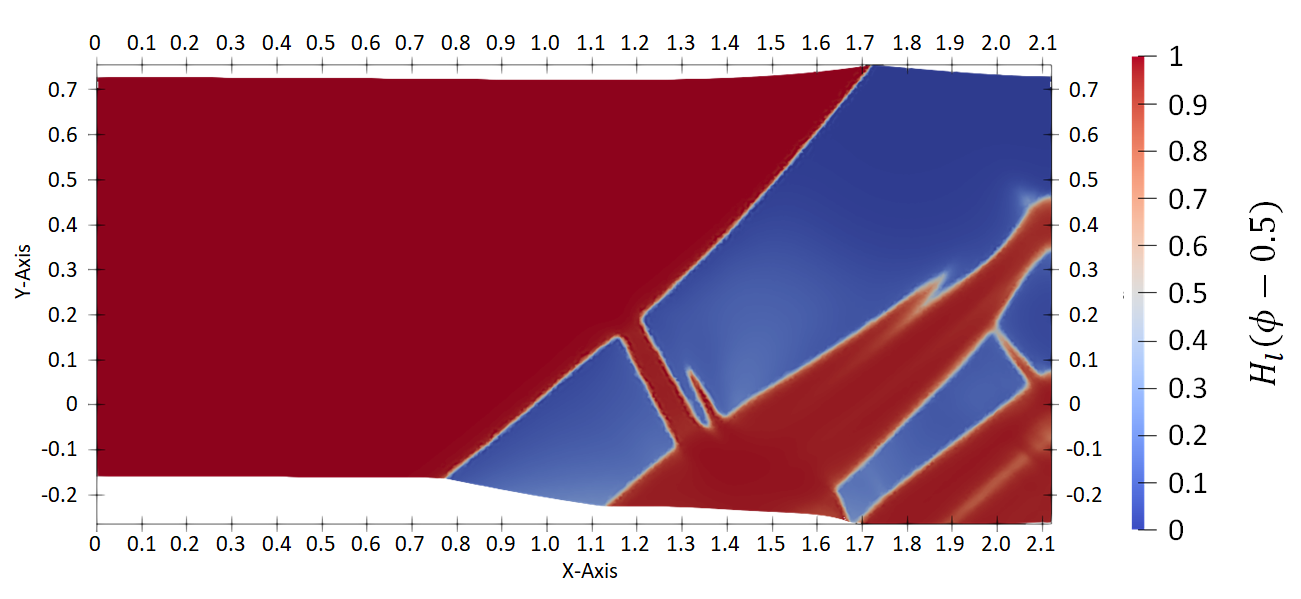}}
    \caption{
        Plots of the deformed configuration overlaid with the phase evolution given by $H_l(\phi-0.5)$. 
        The left column is without nucleation and the right column considers nucleation.
        The first row depicts the initial condition, and the second and third rows depict the evolution in time.
        The imposed loading corresponds to $u_r(t)=\left(\num{1e3} \unit{\milli\metre\per\second}\right) t$.
        We see that accounting for nucleation has a very significant impact on twinning transformation, with large regions well ahead of the moving interface transforming due to nucleation.
        That is, nucleation significantly dominates kinetics.
    }
    \label{consolidation U_Phi with G=0 and G!=0 and eta=0}
\end{figure*}

Figure \ref{consolidation U_Phi with G=0 and G!=0 and eta=0} provides a representative example that highlights the effect of the nucleation term.
The highlight of the calculations shown in Figure \ref{consolidation U_Phi with G=0 and G!=0 and eta=0} is that even at subsonic velocities, it is apparent that accounting for nucleation has a very significant effect and dominates over interface kinetics.
Further, it shows that nonphysical regions in the energy landscape were accessed when nucleation is not accounted for.

We also note that the calculations in Figure \ref{consolidation U_Phi with G=0 and G!=0 and eta=0} do not have viscosity, but the calculations with viscosity ($\eta=\num{1e-9}\unit{s}$) are essentially identical.
This overall conclusion holds for all the calculations described in this section.

\begin{figure*}[htb!]
    \centering
    \subfloat[$E_{11}$]{\includegraphics[scale=0.405]{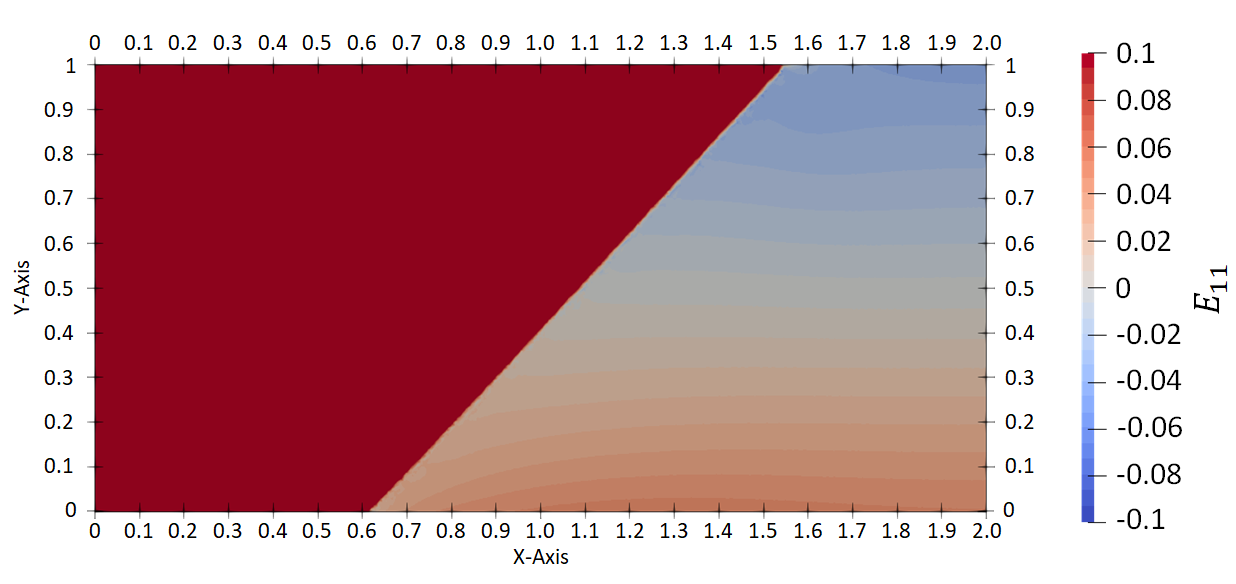}} 
    \hfill
    \subfloat[$E_{22}$]{\includegraphics[scale=0.405]{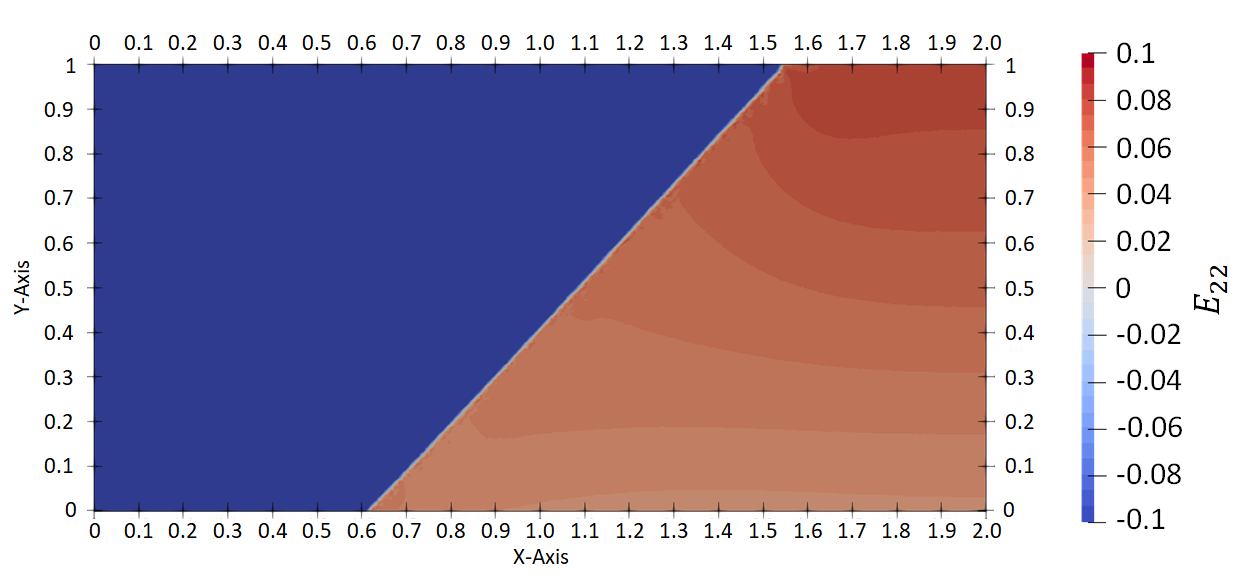}}
    \\
    \subfloat[Line plot of $E_{11}$ and $E_{22}$ along the horizontal centerline.]{\includegraphics[width=0.49\textwidth]{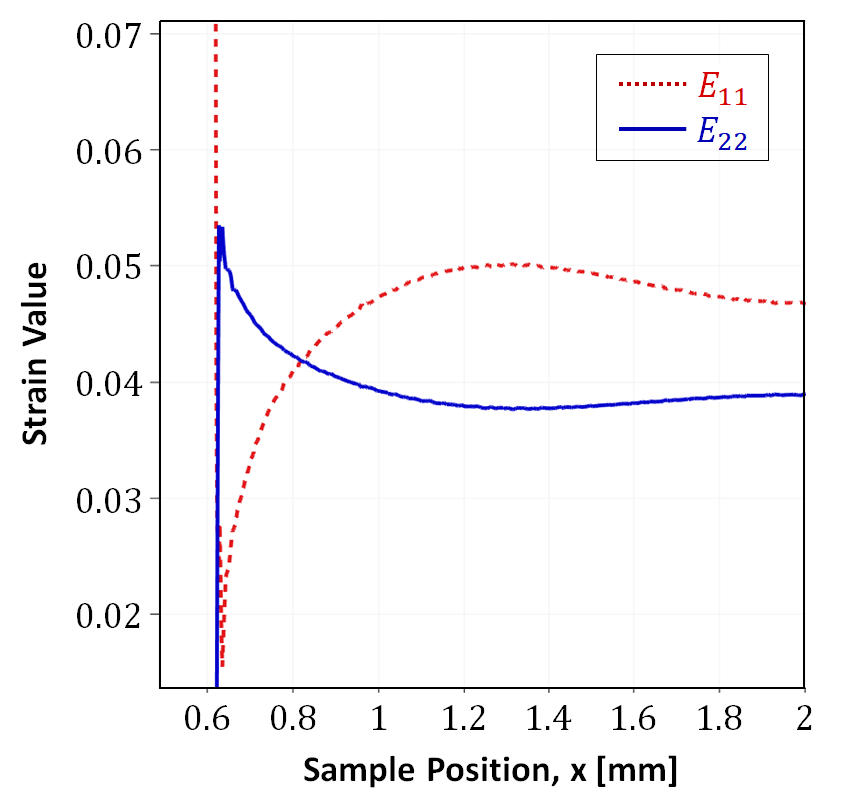}} 
    \caption{
        Plots of $E_{11}$ and $E_{22}$ at $u_r=0.18\unit{mm}$ with a loading rate $u_r(t)=\left(\num{1e3}\unit{\milli\meter\per\second}\right) t$, with nucleation suppressed.
        The system reaches nonphysical regimes in the energy landscape, i.e., the strain and $\phi$ are not consistent with Figure \ref{consolidation G_nucleation illustration}, and this would drive nucleation.
    }
\label{consolidation E22 E11}
\end{figure*}

%%%%%%%%%%%%%%%%
For greater insight into the role of nucleation, Figure \ref{consolidation E22 E11} plots the strain fields in a specimen where the nucleation term is inactive, i.e., the twinning transformation occurs exclusively due to interface motion and not nucleation.
Looking at a time at which the calculation in Figure \ref{consolidation U_Phi with G=0 and G!=0 and eta=0} that allows nucleation had nucleated several new domains, Figure \ref{consolidation E22 E11} shows clearly that $E_{11}>E_{22}$ in the right part of the specimen.
We would thus expect the $\phi=1$ phase to nucleate in that region.

\begin{figure*}[htb!]%
    \centering
    \includegraphics[scale=0.6]{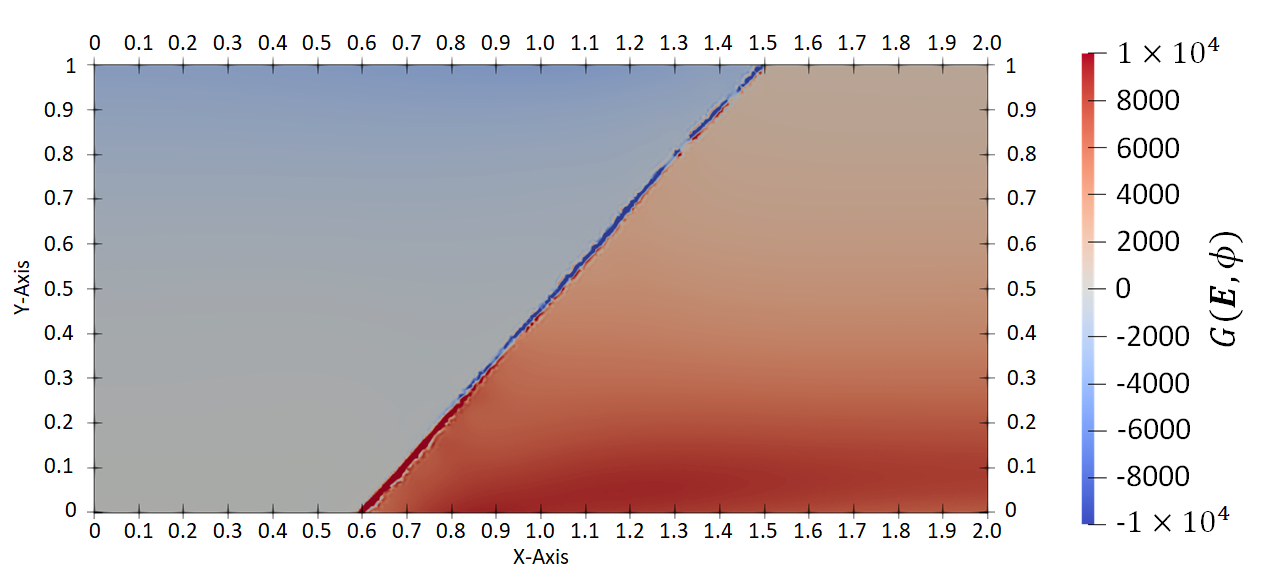}
    \caption{The driving force for nucleation over the specimen, with $u_r=0.04\unit{mm}$ with $G_0=\num{1e5}\unit{s^{-1}}$. }
    \label{consolidation G_nucleation timestep 80}
\end{figure*}

Figure \ref{consolidation G_nucleation timestep 80} shows the driving force for nucleation $G$ as a function of position, at a time when nucleation is active.
We observe the value of $G$ is very high in the lower right part of the specimen.

\begin{figure*}[htb!]%
    \centering
    \subfloat[$u_r=0.18\unit{mm}$, $G_0=0\unit{s^{-1}}$]{\includegraphics[scale=0.405]{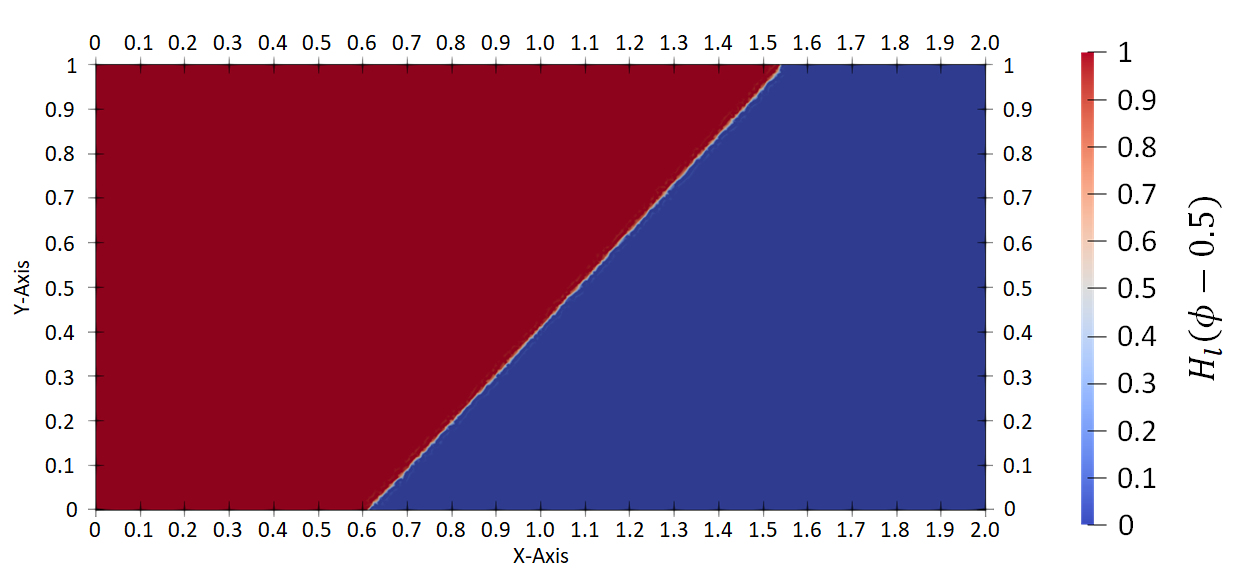}} 
    \hfill
    \subfloat[$u_r=0.18\unit{mm}$, $G_0=\num{1e5}\unit{s^{-1}}$]{\includegraphics[scale=0.405]{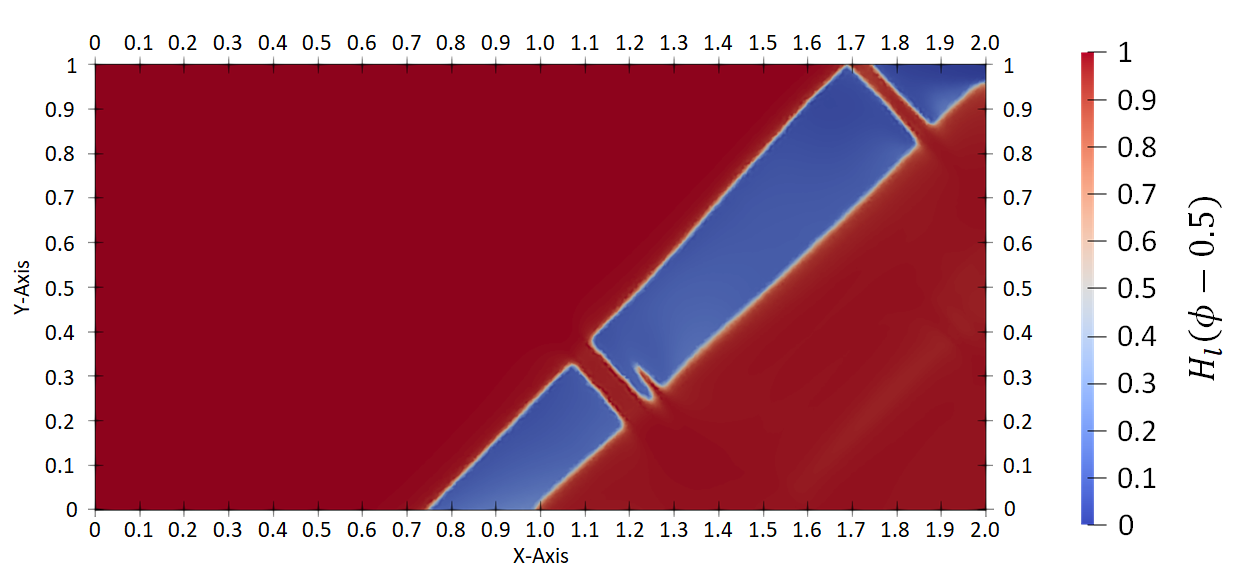}}
    \caption{Plots of $H_l(\phi-0.5)$ for a system (a) without nucleation and (b) with nucleation, with a loading rate $u_r(t)=\left(\num{1e3} \unit{\milli\meter\per\second}\right) t$.
    Comparing against the line plot in Figure \ref{consolidation E22 E11}(c), we see that nucleation has driven twinning through much of the right side of the specimen.
    }
    \label{consolidation withG withoutG 90}
\end{figure*}

Comparing systems with and without nucleation, we observe in Figure \ref{consolidation withG withoutG 90} that the driving force for nucleation $G$ drove the nucleation of the $\phi=1$ phase in the right of the specimen, which corresponds to the position where the value of $G$ was very high several timesteps ago. 
This suggests that the driving force for nucleation has driven the system to a physical region of the energy landscape.

A final point is that when we attempt to drive the interface to propagate faster (via higher loading rates), we encounter pinning --- essentially, very slow evolution --- of the interface where it meets the boundary of the specimen. 
As the twin phases are related by a shear, the driving force for the interface to propagate is related to the shear traction on the plane that is parallel to the interface. 
The boundaries are traction-free on the top and bottom of the domain,  resulting in zero shear stress at the twin interface on the top and bottom boundary.
This effectively pins the interface at the boundaries by causing it to evolve very slowly compared to the interior.
This is further discussed in \cite{agrawal2015dynamic}.

%%%%%%%%%%%%%%%%%%%%%
%%%%%%%%%%%%%%%%%%%%%
%%%%%%%%%%%%%%%%%%%%%
%%%%%%%%%%%%%%%%%%%%%
\section{Anisotropic Kinetics Drives Needle Twin Growth}

%%%%%%%%%%%%%%%%%%%%%
%%%%%%%%%%%%%%%%%%%%%
%%%%%%%%%%%%%%%%%%%%%
%%%%%%%%%%%%%%%%%%%%%

When twins initially nucleate, they grow in a characteristic needle-like geometry, e.g. the experimental observations in \cite{cai-interface}. 
The growth of needle twins was studied using a sharp-interface model using theoretical approaches in several works by P. Rosakis \cite{rosakis-tsai-aniso,rosakis-tsai-jmps2001,rosakis1995dynamic}.
A key finding in that body of work was that anisotropic kinetics is essential to capture needle-like twins.
As discussed in \cite{rosakis-tsai-jmps2001}, the incompatible direction should have more twinning dislocations then the compatible directions to accommodate the deformation, and hence the kinetics will be significantly different for twin motion along compatible and incompatible directions.
This micromechanical effect gives rise to mesoscale anisotropy in the kinetics.

We aim to develop a phase-field description that can capture the phenomenon of needle twin growth.
A key challenge for standard phase-field models is the inability to incorporate anisotropic kinetics.
We develop a phase-field model here that has anisotropic kinetics and show that it is able to capture needle-like twin growth, but also show that incorporating nucleation is essential to prevent nonphysical broadening.
Following \cite{Rosakis-levelset}, we work in the antiplane elastic setting for simplicity.

%%%%%%%%%%%%%%%%%%%%%
%%%%%%%%%%%%%%%%%%%%%
%%%%%%%%%%%%%%%%%%%%%
%%%%%%%%%%%%%%%%%%%%%
\subsection{Model Formulation}

We work in the 2-d anti-plane setting.
The independent space variable is denoted $\bfx = (x_1, x_2)$ in Cartesian coordinates; and the displacement has a single non-zero component in the out-of-plane $3$-direction, denoted $u(\bfx,t)$ where $t$ is the time.
The strain is denoted $\bfeps$, and has only two non-zero components: $\varepsilon_{1} = \half \partial_{x_1} u $ and $ \varepsilon_{2} = \half \partial_{x_2} u$.

We use displacement boundary conditions all around and use this to simulate dynamic loading.
We start with a single phase and have a small nucleus (of size $0.05$) of the second phase near one of the vertical edges to drive the growth of a single needle twin in a convenient location.

%%%%%%%%%%%%%%%%%%%%%
%%%%%%%%%%%%%%%%%%%%%
%%%%%%%%%%%%%%%%%%%%%
%%%%%%%%%%%%%%%%%%%%%
\subsubsection{Elasticity}

We set each phase to have the quadratic form:
\begin{equation}
    W_i(\bfeps) = \half \mu |\bfeps-\bfeps_{0i}|^2
\end{equation}
The material is isotropic in the antiplane setting; see \cite{rosakis-tsai-aniso} for discussion on the anisotropic case.
We use $\mu=1$.

The stress-free strains are $\bfeps_{01} = \mathbf{0}$ and $\bfeps_{02} = \{0.0, 1.0 \}$.
Consequently, the compatible direction is $\bfn = \bfe_2 = \{0,1\}$. 

For viscous dissipation, we use $\eta\dot{\bfeps}$ with $\eta=0.005$.

%%%%%%%%%%%%%%%%%%%%%
%%%%%%%%%%%%%%%%%%%%%
%%%%%%%%%%%%%%%%%%%%%
%%%%%%%%%%%%%%%%%%%%%
\subsubsection{Anisotropic kinetics}
\label{sec:aniso}

We use isotropic kinetics and anisotropic kinetics with the respective forms below:
\begin{align}
 \hat{v} & = \kappa f
 \label{eq:iso_kin}
\\
 \hat{v} & = \kappa f \left(0.1+\frac{|\nabla\phi\cdot\bfa|}{|\nabla\phi|} \right) 
 \label{eq:aniso_kin}
\end{align}
In the anisotropic form, there are two terms in the parentheses: the first is isotropic, and the second introduces anisotropy by its dependence on the interface normal $\nabla\phi$.
It models faster kinetics for portions of the interface that have normal closer to the $\bfa$ direction; we set $\bfa = \bfe_1$ for the numerical calculations.
Further, we use $\kappa=5.0$.

%%%%%%%%%%%%%%%%%%%%%
%%%%%%%%%%%%%%%%%%%%%
%%%%%%%%%%%%%%%%%%%%%
%%%%%%%%%%%%%%%%%%%%%
\subsubsection{Nucleation}

The nucleation term has the following form:
\begin{equation}
        G(\bfeps,\phi) 
        = G_0\left(H_l\left(\phi - 0.5\right)\left(H_l\left(\varepsilon_2 - \bar{\varepsilon}\right)-1 \right) + \left(1-H_l\left(\phi - 0.5\right)\right)H_l\left(\varepsilon_2 - \bar{\varepsilon}\right)\right)
\end{equation}
where $\bar{\varepsilon} = 0.5$.

%%%%%%%%%%%%%%%%%%%%%
%%%%%%%%%%%%%%%%%%%%%
%%%%%%%%%%%%%%%%%%%%%
%%%%%%%%%%%%%%%%%%%%%
\subsection{Comparing Isotropic and Anisotropic Kinetics}

We begin by examining effect of anisotropic kinetics, but do not account for nucleation to enable us to precisely understand their separate roles.

\begin{figure}[htb!]
    \centering
    \subfloat[$t=1$]{\includegraphics[width = 0.4\textwidth]{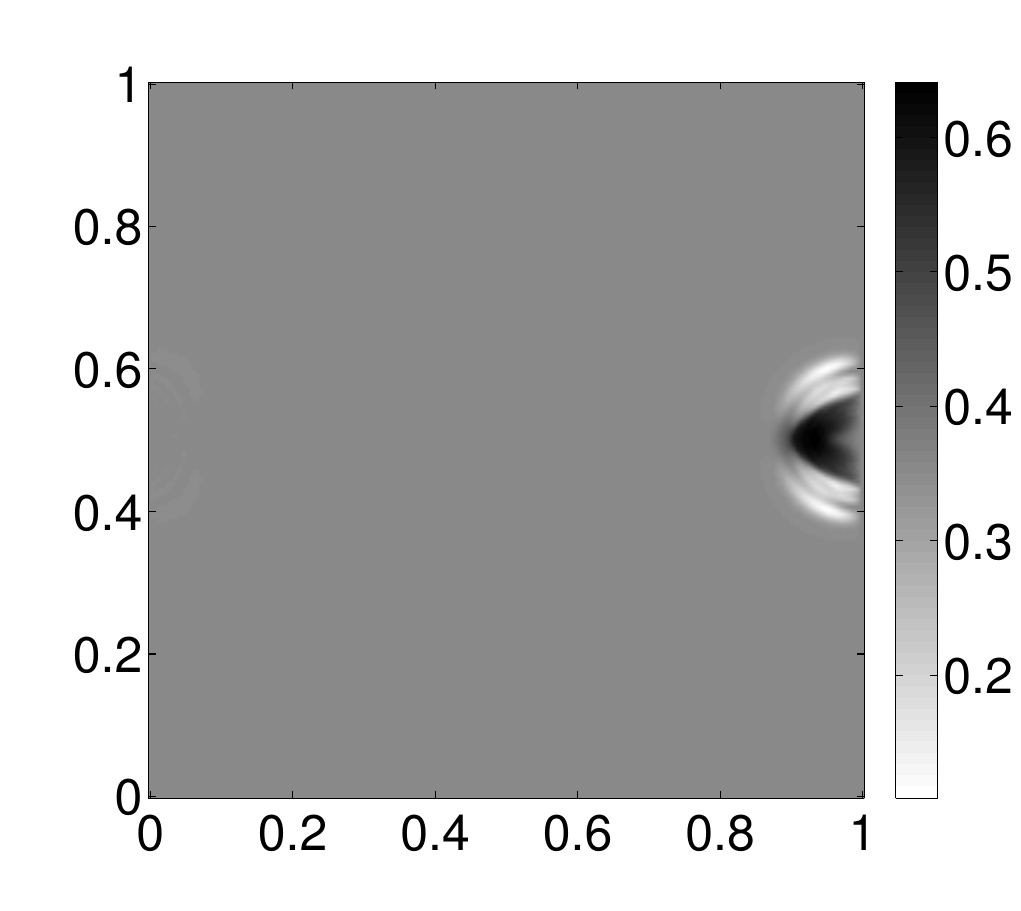}}
    \subfloat[$t=1$]{\includegraphics[width = 0.4\textwidth]{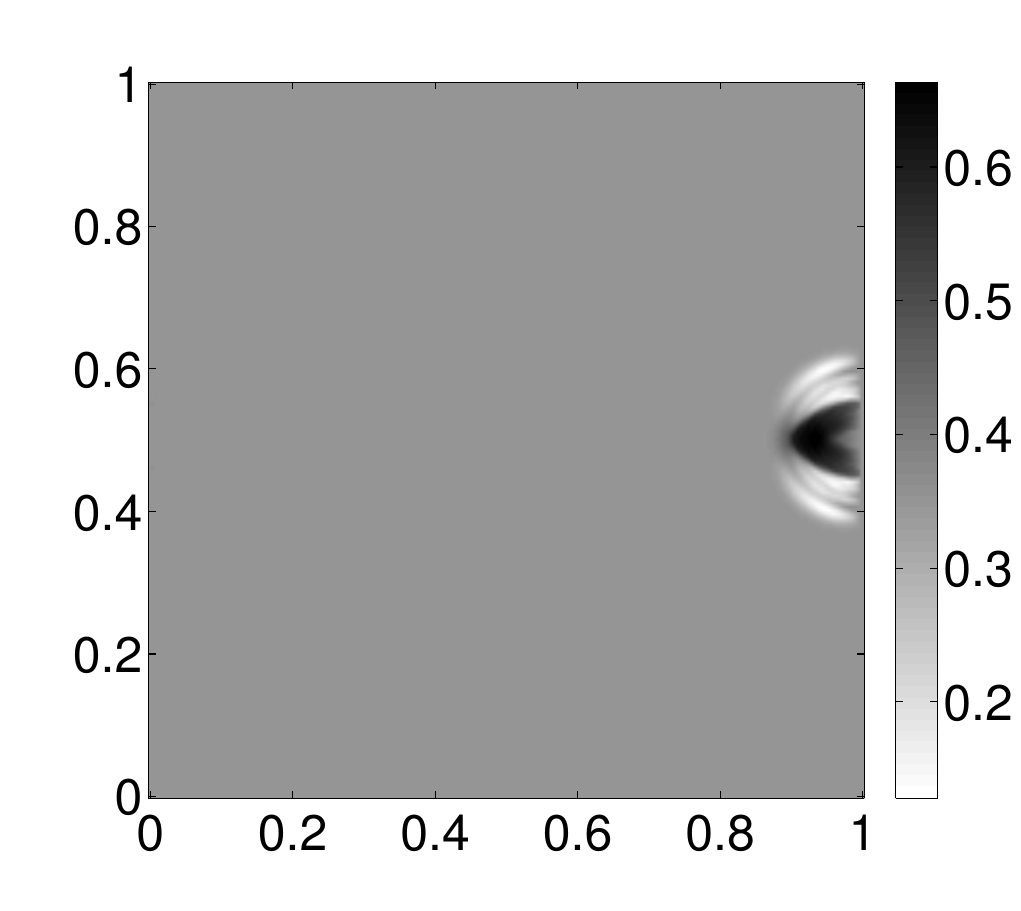}}
    \\
    \subfloat[$t=5$]{\includegraphics[width = 0.4\textwidth]{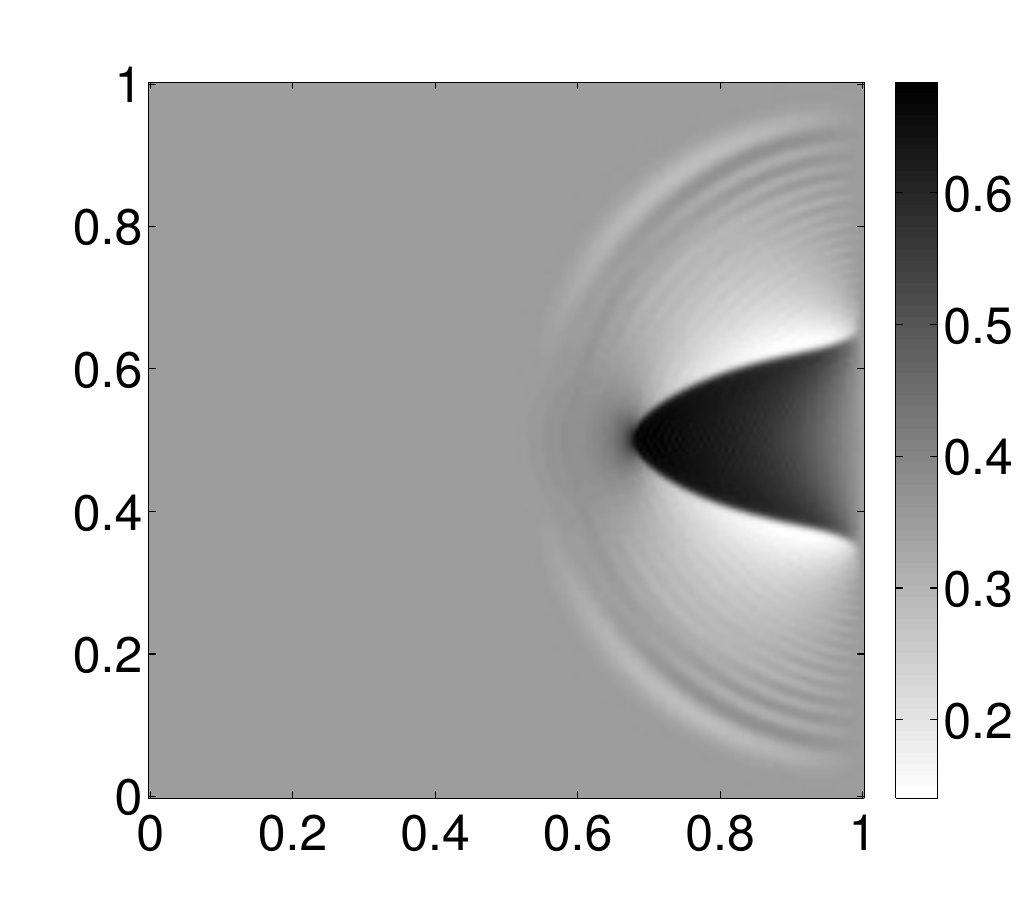}}
    \subfloat[$t=5$]{\includegraphics[width = 0.4\textwidth]{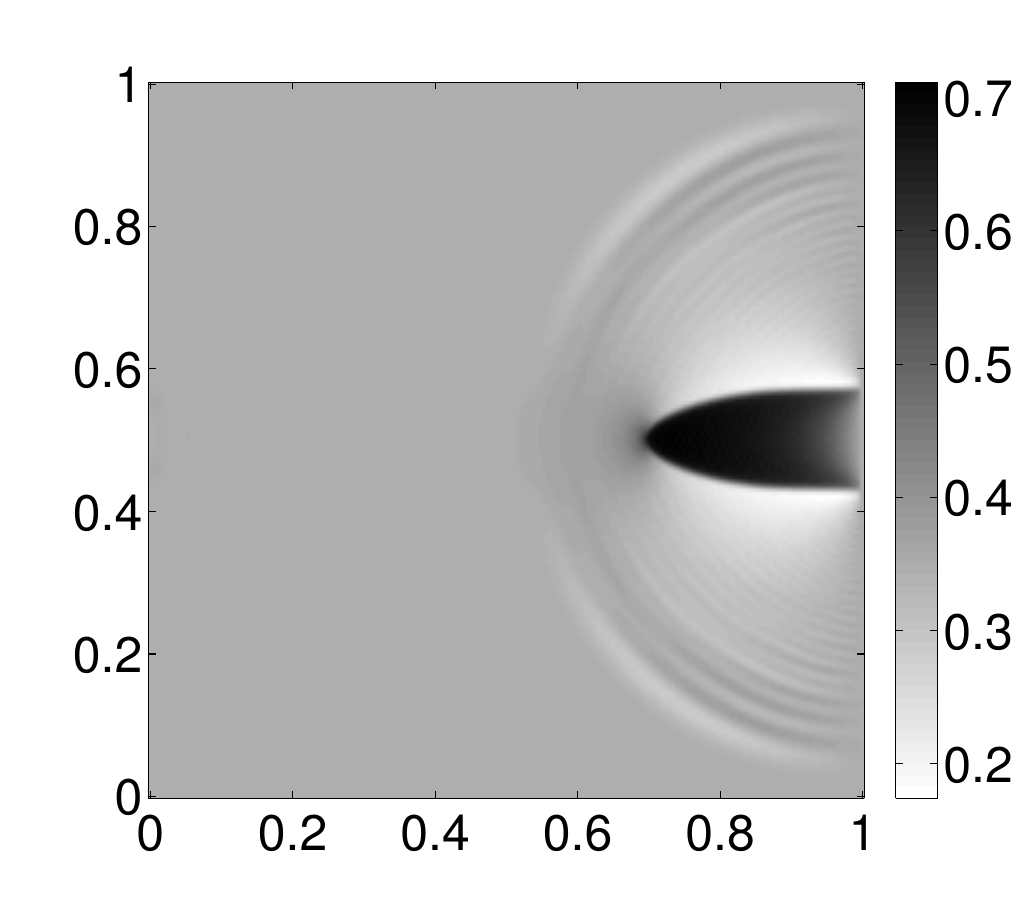}}
    \\
    \subfloat[$t=10$]{\includegraphics[width = 0.4\textwidth]{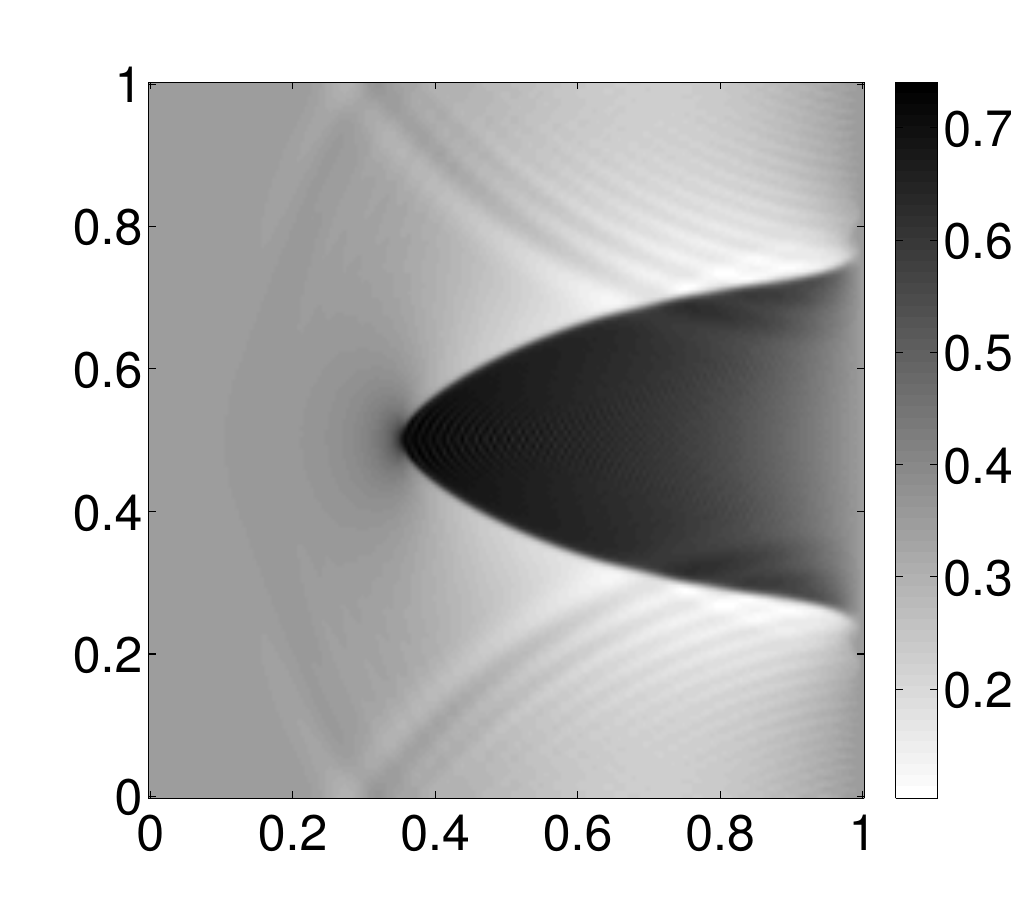}}
    \subfloat[$t=10$]{\includegraphics[width = 0.4\textwidth]{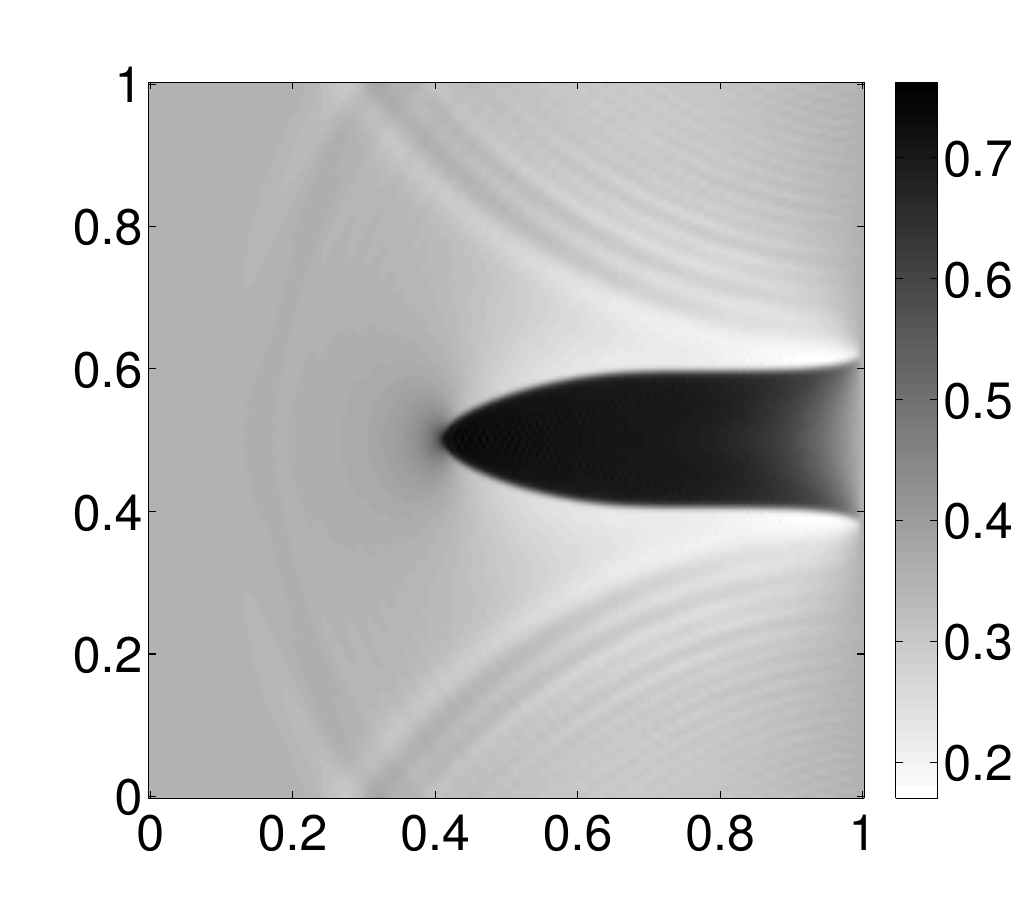}}
    \caption{Time evolution of $\varepsilon_2$ ($t=1,5,10$). 
    The left and right columns are respectively isotropic and anisotropic kinetics, with faster propagation when the local interface normal is closer to being aligned with $\bfe_1$.
    }
    \label{fig:kin-horiz}
\end{figure}

Figure \ref{fig:kin-horiz} shows the evolution of the twin needle in time, comparing isotropic kinetics to anisotropic kinetics.
We see that anisotropy is essential to obtain needle-like twin growth, consistent with the work of \cite{rosakis-tsai-jmps2001}.
However, we also notice that the twin slowly gets wider in the thin direction beginning at the base.

%%%%%%%%%%%%%%%%%%%%%%%%%%%%%%%%%%%%%%%%%%%%%%%%%%%%%%%%%%%%%%%
%%%%%%%%%%%%%%%%%%%%%%%%%%%%%%%%%%%%%%%%%%%%%%%%%%%%%%%%%%%%%%%
%%%%%%%%%%%%%%%%%%%%%%%%%%%%%%%%%%%%%%%%%%%%%%%%%%%%%%%%%%%%%%%
%%%%%%%%%%%%%%%%%%%%%%%%%%%%%%%%%%%%%%%%%%%%%%%%%%%%%%%%%%%%%%%
\subsection{Effect of Nucleation and Viscous Stress}

\begin{figure}[htb!]
    \centering
    \includegraphics[width = 0.8\textwidth]{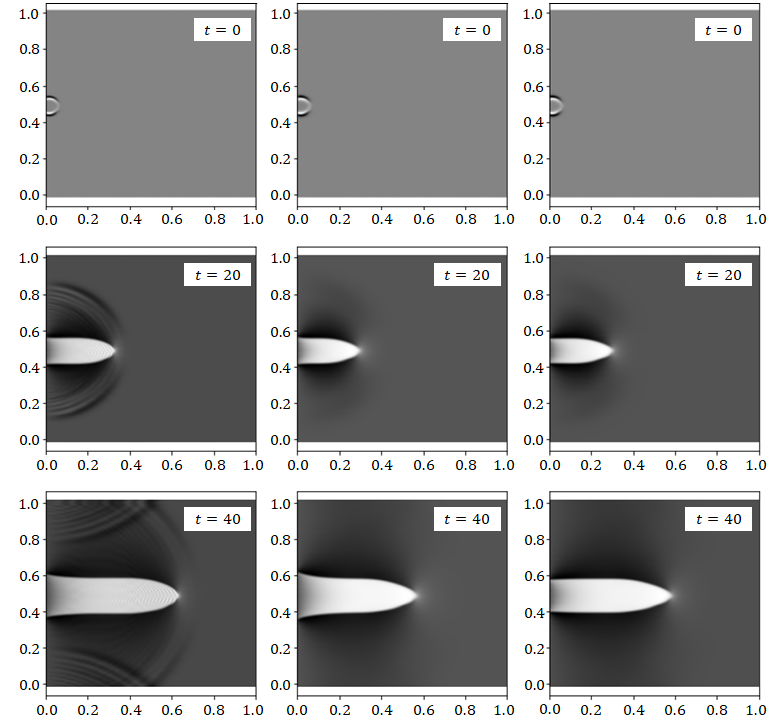}
    \caption{Time evolution of $\varepsilon_2$ ($t=0, 20, 40$).
    The left column accounts only for anisotropic kinetics; the middle column includes also viscous stress; and the right column with anisotropic kinetics, viscous stress, and nucleation.
    We highlight that including viscous stress and nucleation appears to suppress the nonphysical broadening of the twin as it grows.
    }
    \label{fig:old_vs_augm}
\end{figure}

Building on the observation that anisotropic kinetics appears essential to predict realistic needle twins, we next consider the effect of further accounting for viscous stress and nucleation.
Figure \ref{fig:old_vs_augm} compares the effect of accounting for nucleation and viscous stresses in addition to anisotropic kinetics.
We see that viscous stresses attenuate the leading elastic waves but do not significantly affect the needle twin microstructure.
In contrast, the nucleation term serves the important purpose of suppressing the nonphysical broadening of the twin.

\begin{figure}[htb!]
    \centering
    \includegraphics[width = \textwidth]{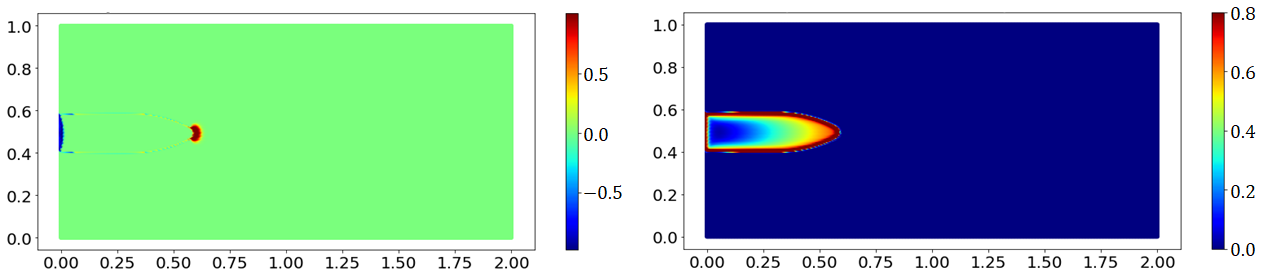}
    \caption{The driving force for nucleation (left) and kinetics (right) at $t=40$. Both driving forces have comparable magnitudes, and the nucleation driving force acts at the base of the needle twin to prevent nonphysical broadening.}
    \label{fig:highKin_Gfix_dissip}
\end{figure}

We further plot the driving forces for kinetics and nucleation in Figure \ref{fig:highKin_Gfix_dissip}.
First, we notice that both driving forces are of comparable magnitude.
Second, we highlight that the nucleation term is active at the growing tip as expected, but also at the base of the needle twin where it acts to prevent nonphysical twin broadening.

%%%%%%%%%%%%%%%%%%%%%%%%%%%%%%%%%%%%%%%%%%%%%%%%%%%%%%%%%%%%%%%
%%%%%%%%%%%%%%%%%%%%%%%%%%%%%%%%%%%%%%%%%%%%%%%%%%%%%%%%%%%%%%%
%%%%%%%%%%%%%%%%%%%%%%%%%%%%%%%%%%%%%%%%%%%%%%%%%%%%%%%%%%%%%%%
%%%%%%%%%%%%%%%%%%%%%%%%%%%%%%%%%%%%%%%%%%%%%%%%%%%%%%%%%%%%%%%
\subsection{Supersonic Twin Growth}

\begin{figure}[htb!]
\begin{center}
    \includegraphics[width = \textwidth]{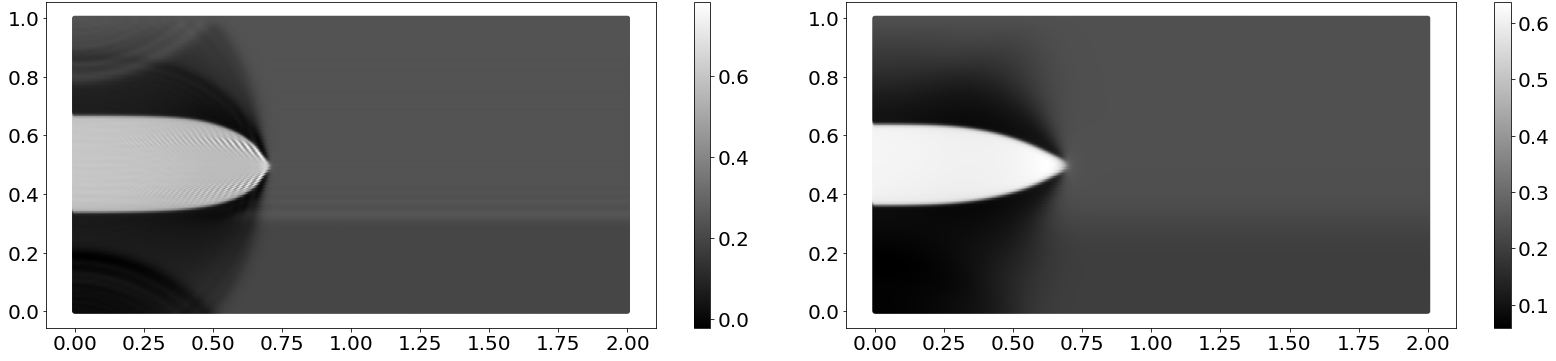}
    \caption{Supersonic twin growth needles visualized through plotting $\varepsilon_2$ ($t=40$). The left figure accounts only for anisotropic kinetics and predicts a somewhat-rounded twin growing at Mach $1.06$.
    The right figure accounts further for nucleation and viscous stress, and predicts a more needle-like twin that grows slightly faster at Mach $1.13$.}
    \label{fig:supsonic_noaug_vs_aug_F}
\end{center}
\end{figure}

We next look at the role of nucleation and viscous stress in supersonic twinning.
To prevent the use of overly large stresses, we achieve supersonic twinning by increasing the kinetic coefficient $\kappa$ by a factor of $10$.
Figure \ref{fig:supsonic_noaug_vs_aug_F} compares the twin microstructure in a model with only anisotropic kinetics to a model that also accounts for nucleation and viscous stress.
As can be inferred from the Mach cone structure, both twins grow supersonically, with the model with viscous damping showing faster twin growth.

\begin{figure}[htb!]
    \begin{center}
    \includegraphics[width = \textwidth]{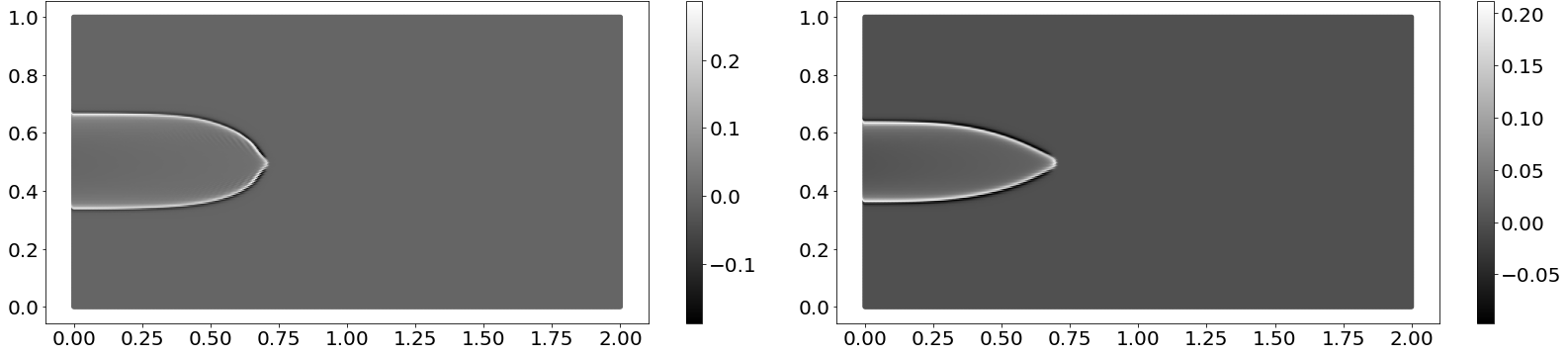}
    \caption{The kinetic driving force in the model without (left) and with (right) nucleation and viscous stress.}
    \label{fig:eq_mod_dforce}
    \end{center}
\end{figure}

We next plot the kinetic driving force for both models in Figure \ref{fig:eq_mod_dforce}, and notice that there are no significant differences.
As expected in our class of models, the driving force is localized around the moving interface \cite{chua2022phase,agrawal2015dynamic}.

\begin{figure}[htb!]
    \begin{center}
	\includegraphics[width = \textwidth]{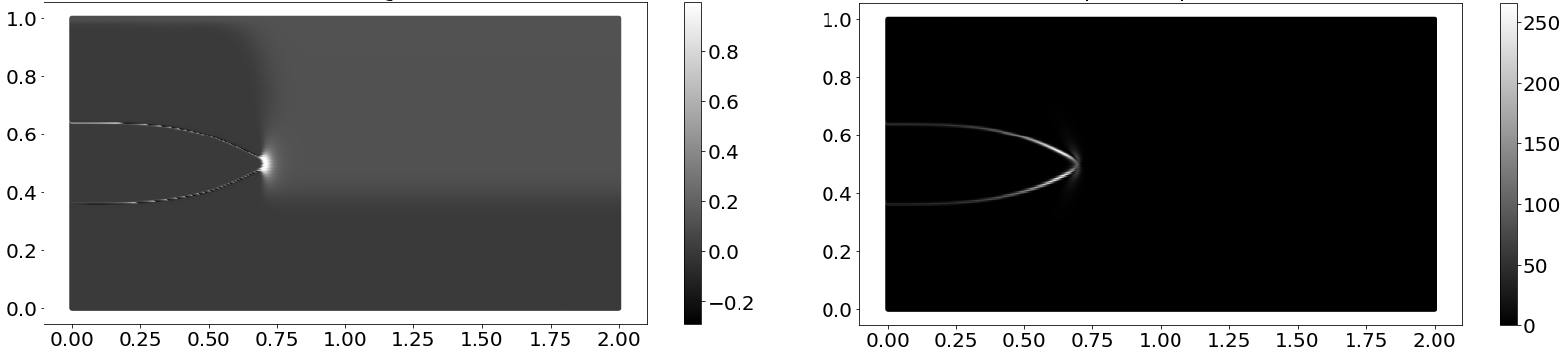}
    \caption{Left: the driving force for nucleation, which has magnitude comparable to the kinetic driving force (Fig. \ref{fig:eq_mod_dforce}). Right: the viscous stress. Both are highly localized at the growing tip of the needle twin.}
    \label{fig:eq_mod_pseudoG}
    \end{center}
\end{figure}

Figure \ref{fig:eq_mod_pseudoG} shows the nucleation driving force and the viscous stress.
We notice that both are highly localized at the tip of the growing twin, and that the largest value of the nucleation driving force is comparable to the kinetic driving force.

\begin{remark}[Supersonic Twinning with Unequal Moduli]
    We have performed similar supersonic twinning calculations for the case where the moduli are unequal with the ratio $1:4$.
    When the twin propagates into the soft phase, the results are qualitatively similar to the discussion above.
    When the twin propagates into the stiff phase, the needle twin appears unstable and splits up into a more complex microstructure.
\end{remark}

%%%%%%%%%%%%%%%%%%%%%
%%%%%%%%%%%%%%%%%%%%%
%%%%%%%%%%%%%%%%%%%%%
%%%%%%%%%%%%%%%%%%%%%
\section{Discussion}

We have used a recent phase-field modeling framework proposed in \cite{chua2022phase,agrawal2015dynamic,agrawal2015dynamic-2} to study the interplay between kinetics and nucleation in the dynamic evolution of twins.
The ability of this framework to explicitly and transparently specify nucleation and kinetic behavior is essential to model phenomena such as needle twin growth.
For instance, anisotropic interface kinetics is essential to model needle twin growth, but it is unclear 
how --- or if it is even possible --- to specify anisotropic kinetics in standard phase-field approaches or alternatives such as peridynamics.

For the problems studied in this paper, we find that viscous stresses do not play an important role.
This is a significant difference compared to studies in other settings, e.g., \cite{faye2017spherical,chua2022phase}.
In contrast to those works, the process of twinning in higher dimensions is dominated by geometric compatibility, which leads to this significant difference.

%%%%%%%%%%%%%%%%%%%%%
%%%%%%%%%%%%%%%%%%%%%
%%%%%%%%%%%%%%%%%%%%%
%%%%%%%%%%%%%%%%%%%%%
\paragraph*{Software and Data Availability.}

The code developed for this work and the associated data are available at \\ \url{https://github.com/janelchua/Phase-field_Twin-interface}

\paragraph*{Acknowledgments.}
    We thank NSF (2108784, 2012259) and ARO (MURI W911NF-19-1-0245) for support; NSF for XSEDE computing resources provided by Pittsburgh Supercomputing Center; Maryam Khodadad for reviewing the code; and Tony Rollett and Phoebus Rosakis for useful discussions.

%%%%%%%%%%%%%%%%%%%%%
%%%%%%%%%%%%%%%%%%%%%
%%%%%%%%%%%%%%%%%%%%%
%%%%%%%%%%%%%%%%%%%%%

% References

\end{document}